\documentclass[aps,prl,twocolumn]{revtex4-1}
\usepackage{graphicx,epsfig,epstopdf}
\usepackage{amsmath}
\usepackage{subfigure}
\usepackage{mathrsfs}
\usepackage{bm}
\usepackage{times}
\begin{document}

\title{Superadiabatic Holonomic Quantum Computation in Cavity QED}

\author{Bao-Jie Liu}
\affiliation{Guangdong Provincial Key Laboratory of Quantum Engineering and Quantum Materials and School of Physics\\ and
Telecommunication Engineering, South China Normal University, Guangzhou 510006, China}

\author{Zhen-Hua Huang}
\affiliation{Guangdong Provincial Key Laboratory of Quantum Engineering and Quantum Materials and School of Physics\\ and Telecommunication Engineering, South China Normal University, Guangzhou 510006, China}

\author{Zheng-Yuan Xue}\email{zyxue@scnu.edu.cn}
\affiliation{Guangdong Provincial Key Laboratory of Quantum Engineering and Quantum Materials and School of Physics\\ and Telecommunication Engineering, South China Normal University, Guangzhou 510006, China}

\author{Xin-Ding Zhang}\email{xdzhang@scnu.edu.cn}
\affiliation{Guangdong Provincial Key Laboratory of Quantum Engineering and Quantum Materials and School of Physics\\ and Telecommunication Engineering, South China Normal University, Guangzhou 510006, China}


\date{\today}

\begin{abstract}
Adiabatic quantum control is a powerful tool for quantum engineering and a key component in some quantum computation models, where accurate control over the timing of the involved pulses is  not needed. However, the adiabatic condition requires that the process should be very slow and thus limits its application in quantum computation, where quantum gates are preferred to be fast due to the limited coherent times of the quantum systems. Here, we propose a feasible scheme to implement universal holonomic quantum computation based on non-Abelian geometric phases with superadiabatic quantum control, where the adiabatic manipulation is sped up while retaining its robustness against errors in the timing control. Consolidating the advantages of both strategies, our proposal is thus both robust and fast. The quantum cavity QED system is adopted as a typical example to illustrate the merits, where the proposed scheme can be realized in a tripod configuration by appropriately controlling the pulse shapes and their relative strength. To demonstrate the distinct performance of our proposal, we also compare our scheme with the conventional adiabatic strategy.
\end{abstract}



\maketitle

A practical quantum computer must be capable of implementing high-fidelity quantum gates on a scalable array of quantum qubits. In the presence of noises, the physical realization of quantum computation is posed daunting challenges.
Geometric phases \cite{b1,b2,b3}, which possess intrinsic noise-tolerant features, are promising for performing robust quantum computations \cite{gqc,b4,b5,adiabatic1,b6,Solinas}. In particular, quantum  holonomies  \cite{b3}, i.e., non-Abelian geometric phases, naturally lead to universal quantum computation due to their non-commutativity. Although quantum gates based on adiabatic holonomies have already been proposed \cite{adiabatic1,b6,adiabatic2,adiabatic3,xdzhang,adiabatic4,adiabatic5,adiabatic6,Kumar}, the adiabatic evolution reduces the gate speed, and thus, decoherence effects will introduce unacceptable errors \cite{b4,b5}. One possible way out of this dilemma is the so-called transitionless quantum driving protocol \cite{superadiabatic1,superadiabatic2,superadiabatic3,superadiabatic4}, where the nonadiabatic transition during the evolution is suppressed by an additional driving field,   even when the adiabatic condition is not met. Therefore, great efforts have been made in this field \cite{xchen,ga,mr,b17,b18,b15,b16,b19}. In particular, for the Abelian geometric phase case, this protocol is further simplified by only modifying the driving fields in the adiabatic case \cite{xydu,dress,dressexp}.

In this letter, we generalize the case to the non-Abelian case, i.e., the implementation of superadiabatic holonomic quantum computation with only the modification of the pulse shape of the driving fields in the adiabatic case, and thus greatly reduce the experimental difficulties. In comparison with standard holonomic quantum computation, the holonomies obtained in our approach tend asymptotically to those of the adiabatic approach in the long run-time limit and thus might open up a new horizon for realizing a practical quantum computer. We illustrate our idea in a cavity quantum electrodynamics (C-QED) system \cite{UQ}, which avoids the adiabatic condition while maintaining the advantage of robustness.

We present our scheme using the basic factors for the experimental realization of a two-atom C-QED with cesium atoms \cite{Atom}, which have been cooled and trapped in a small optical cavity in the strong coupling regime. We defined that the state $\left| {e} \right\rangle$ corresponds to the F=4, m=3 hyperfine state of the $6^2P_{1/2}$ electronic excited state; the state $\left| {2} \right\rangle$ corresponds to the F=4, m=3 hyperfine state of the $6^2S_{1/2}$ electronic ground state; the state $\left| {1} \right\rangle$ corresponds to the F=3, m=2 hyperfine state of the $6^2S_{1/2}$ electronic ground state; and the state $\left| {0} \right\rangle$ corresponds to the F=3, m=4 hyperfine state of the $6^2S_{1/2}$ electronic ground state. For one four-level atom  with three ground states $\left| {i} \right\rangle$ ($i=0, 1, 2$) and an excited state $\left| {e} \right\rangle$, a sketch of the level structure is shown in Fig.1(a). The atomic transition from $\left| {i} \right\rangle$ to $\left| {e} \right\rangle$ is driven resonantly through a classical laser field with the time-dependent Rabi frequency ${\Omega_i}(t)$. Under the rotating-wave approximation (RWA), the interaction Hamiltonian for this system reads ($\hbar=1$ herein)
\begin{equation}\label{1}
H(t)=\sum_{i=0}^{2}\Omega_{i}|e\rangle\langle i|+\text{h.c.},
\end{equation}
where the amplitudes of the driving fields are parameterized  as
$\Omega_{0}(t) = \Omega(t)\sin\theta(t)\sin\varphi$, 
$\Omega_{1}(t) = \Omega(t)\sin\theta(t)\cos\varphi$, 
and $\Omega_{2}(t) = \Omega(t)\cos\theta(t) e^{-i\phi}$
with $\tan\varphi=\Omega_{0}(t)/ \Omega_{1}(t)$ and
$\tan\theta(t)=\sqrt{\Omega^{2}_{0}(t)+\Omega^{2}_{1}(t)}/ \Omega_{2}(t)$. With these choices, the instantaneous eigenstates are \cite{b12}
\begin{eqnarray}
\left| {{d_1}} \right\rangle  &=&  \cos \theta \left( t \right)|\psi\rangle - \sin \theta \left( t \right){e^{i\phi }}\left| 2 \right\rangle,  \notag \\
\left| {{d_2}} \right\rangle  &=& \cos \varphi \left| 0 \right\rangle  - \sin \varphi \left| 1 \right\rangle, \notag \\
\left|  +  \right\rangle  &=& \frac{1}{{\sqrt 2 }}[\sin \theta ( t ) |\psi\rangle
+ \cos \theta \left( t \right){e^{i\phi }}\left| 2 \right\rangle
- \left| e \right\rangle],  \notag \\
| -\rangle  &=& \frac{1}{{\sqrt 2 }}[\sin \theta \left( t \right) |\psi\rangle
+ \cos \theta \left( t \right){e^{i\phi }}\left| 2 \right\rangle
+ \left| e \right\rangle ]
\end{eqnarray}
with $|\psi\rangle=(\sin \varphi \left| 0 \right\rangle  + \cos \varphi \left| 1 \right\rangle )$ . The corresponding four eigenvalues are $E_{d_{j}}=0$ ($j=1, 2$) and $E_{\pm}=\pm\hbar\Omega/2$, with $\Omega=\sqrt{\Omega^{2}_{0}(t)+\Omega^{2}_{1}(t)+\Omega^{2}_{2}(t)}$.

\begin{figure}[tbp]
\centering
\includegraphics[width=0.98\columnwidth]{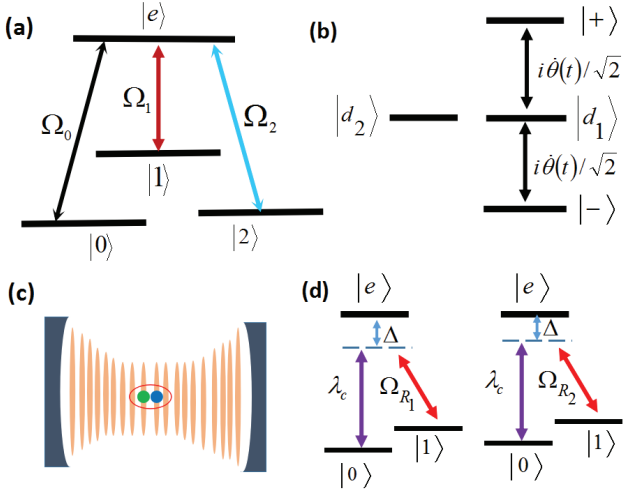}
\caption{Illustation of the proposed scheme. (a) The level structure and coupling configuration for single-qubit operations, where driven pulses with amplitudes $\Omega_{i}(t)$ couple $|i\rangle$  resonantly to $|e\rangle$. (b) Schematic of the level structure in the dressed state space, where the dark state $|d_{2}\rangle$ is decoupled from $|+\rangle$, $|-\rangle$ and $|d_{1}\rangle$ when $\varphi$ is kept constant.  (c) Illustration of a two-qubit gate with two atoms coupled to a cavity in a two-photon Raman resonant way, as shown in (d).} \label{fig1}
\end{figure}

It is noted that the $|d_{2}\rangle$ is decoupled from the other states when $\varphi$ is kept constant; see Fig. \ref{fig1}(b). Therefore, the quantum dynamics is governed by a three-level system Hamiltonian in the basis $\{|+\rangle, |d_{1}\rangle, |-\rangle\}$ as
\begin{equation}\label{3level}
    H_{ad}(t)=\frac{\Omega(t)}{2}M_{z}+\dot{\theta}(t)M_{y},\\
\end{equation}
where $M_{z}=|+\rangle\langle+|-|-\rangle\langle-|$, $M_{x}=(|-\rangle-|+\rangle)\langle d_{1}|/\sqrt{2}+ \text{h.c.}$, and $M_{y}=i(|-\rangle+|+\rangle)\langle d_{1}|/\sqrt{2})+ \text{h.c.}$
are Pauli matrixes for spin 1 systems. The second term of the Hamiltonian in Eq. (\ref{3level}) corresponds to the nonadiabatic transitions between the dark state $|d_{1}\rangle$ and the two bright states $|+\rangle$ and $|-\rangle$ when the adiabatic condition is not met. To correct the nonadiabatic errors, we look for a correction Hamiltonian $H_{C}(t)$ by generalizing the modified pulse method \cite{dress,b15} to our non-Abelian case, i.e., modifying the amplitudes of the pulses so that they cancel the unwanted off-diagonal elements in Eq. (\ref{3level}). The modified pulses are \cite{sm}
\begin{eqnarray}\label{correction}
  \Omega^{'}_{0}(t) &=& \Omega^{'}(t)\sin\theta^{'}(t)\sin\varphi, \nonumber\\
  \Omega^{'}_{1}(t) &=& \Omega^{'}(t)\sin\theta^{'}(t)\cos\varphi, \nonumber\\
  \Omega^{'}_{a}(t) &=& \Omega^{'}(t)\cos\theta^{'}(t)e^{-i\phi},
\end{eqnarray}
where
\begin{eqnarray}\label{modified}
  \theta^{'}(t) &=& \theta(t)-\arctan[\dot{\mu}(t)/\Omega(t)], \nonumber\\
  \Omega^{'}(t) &=& \sqrt{\Omega^{2}(t)+\dot{\mu}^{2}(t)}, \nonumber\\
         \mu(t) &=&  -\arctan[\dot{\theta}(t)/\Omega(t)].
\end{eqnarray}
We now show how to build up a universal single-qubit gate using the above superadiabatic protocol in a cyclic evolution. Under the cyclic evolution, the dynamics is denoted by a unitary evolution
$U=P\exp[-i\int^{2T}_{0}d\lambda H(\lambda)]$,  with $P$ being the time-ordering operator. Here, we consider that the operation procedure is divided into two steps. During the first step, from time $t=0$ to $t=T$,  the states evolve from $|d_{1}(0)\rangle=|\psi\rangle$ to $|d_{1}(T)\rangle=|2\rangle$.  The coupling is governed by the Hamiltonian Eq. (\ref{1}), with the modified parameters given in Eq. (\ref{correction}) and a constant phase $\phi=\phi_{1}$. During the second step, from time $t=T$ to $t=2T$, and for the states from $|d_{1}(T)\rangle=|2\rangle$ to $|d_{1}(2T)\rangle=|\psi\rangle$, the coupling is also governed by the Hamiltonian Eq. (\ref{1}) but with a different constant phase $\phi=\phi_{2}$. Considering two stages, a closed path in the parameter space is formed. The solid angle enclosed by the closed path is evidently $2(\phi_{2}-\phi_{1})$, and thus, the geometric phase acquired is simply $\gamma_{1}=\phi_{2}-\phi_{1}$. Initially, the two dark states are $|d_{1}(0)\rangle=|\psi\rangle$ and $|d_{2}(0)\rangle=\sin\varphi|0\rangle+\cos\varphi|1\rangle$.
Therefore, the unitary on the subspace $\{|d_{1}\rangle,|d_{2}\rangle\}$ is
\begin{equation}
U_{1}=\left(
  \begin{array}{cc}
   e^{i\gamma_{1}} & 0\\
   0 & 1
  \end{array}
\right).
\end{equation}
In the computational space spanned by $\{|0\rangle, |1\rangle\}$, we obtain
\begin{equation}
U_{1}=\left(
  \begin{array}{cc}
   \cos^{2}{\varphi}+e^{i\gamma_{1}} \sin^{2}{\varphi} & \cos{\varphi}\sin{\varphi}(e^{i\gamma_{1}}-1)\\
   \cos{\varphi}\sin{\varphi}(e^{i\gamma_{1}}-1) & \sin^{2}{\varphi}+e^{i\gamma_{1}} \cos^{2}{\varphi}
  \end{array}
\right),
\end{equation}
which can generate a universal set of single-qubit gates, i.e., any desired single-qubit gate can be realized via the proper choice of $\varphi$ and $\gamma_{1}$. For example, when we set $\varphi=\pi /4$ and $\gamma_{1}=\pi$,
a NOT gate is implemented. On the other hand, one can set $\varphi=\pi /8$ and $\gamma_{1}=\pi$ to implement a Hadamard gate.

\begin{figure}[tbp]
\begin{center}
\includegraphics[width=0.98\columnwidth]{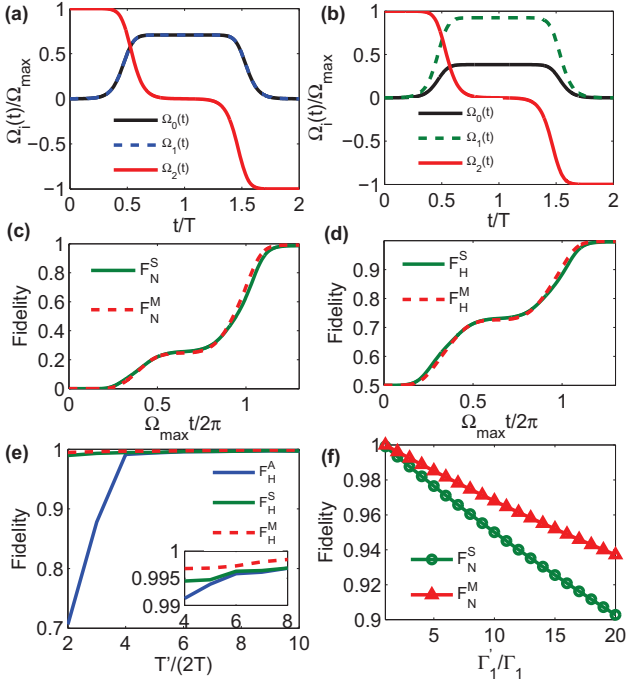}
\caption{Illustration of the performance of the proposed single gates. The original zero-detuned laser pulses with frequency $\Omega_{i}(t)$ for the adiabatic single-qubit NOT and Hadamard gates are sketched in (a) and (b) with the gate times as $2T$. Qubit-state fidelity dynamics of the NOT gate (c) and the Hadamard gate (d) as a function of the dimensionless time $\Omega_{max}t/2\pi$. (e) The fidelity of the Hadamard gates for different operation times $T^{'}$ (in units of 2T). (f) The fidelity of the superadiabatic and modified superadiabatic NOT gates for different rates of the atom $\Gamma^{'}_{1}$ (in units of $\Gamma_{1}$). }
\label{fig2}
\end{center}
\end{figure}

For demonstration purposes, we illustrate such an implementation based on the above C-QED system.  We consider that the Vitanov-style pulses \cite{Vitanov} are
\begin{eqnarray}
 \theta=\left\{
\begin{array}{ll}
\frac{\pi}{2+2e^{-(t-T/2)/\tau}},\ \ (0\leq t<T)\\
\frac{\pi}{2}- \frac{\pi} {2+2e^{-(t-3T/2)/\tau}},\ \ (T\leq t<2T)\\
\end{array}
\right.\notag\\
 \quad \Omega(t)=\Omega_{max},\ \ \ \ \ \ \ \ \ \ \ \ \ \ \ \ \ \ \ \  \ \ \ \ \ \ \ \ \ \ \ \ \ \ \ \ \ \ \  \ \ \ \ \ \ \ \ \ \ \ \ 
\end{eqnarray}
where the time $\tau$ controls the effective duration of the protocol. To simulate pulses with a finite duration, we choose $T=10\tau$ such that $\Omega_{2}(0)=\Omega_{j}(T)< 0.1\Omega_{max}$. As shown in Fig. \ref{fig2}(a) and \ref{fig2}(b), we plot the pulse shape of both the original and modified pulse shape for the  NOT and Hadamard gates, respectively.

The decoherence process in the Cavity QED system in the described implementation is unavoidable, and understanding its effects is crucial for our scheme. The performance of the gates can be evaluated by considering the influence of dissipation using the Markovian master equation in Lindblad form
\begin{equation}\label{10}
\dot{\rho}=-i[H(t),\rho]+\frac{\kappa}{2}L(a)
    +\sum_i \left[\frac{\Gamma_{1}}{2}L(A_i)+\frac{\Gamma_{2}}{2}L(B_i)\right]
\end{equation}
where $\rho$ is the density matrix of the considered system, $A_i=|e\rangle\langle i|$, $B_i=|e\rangle\langle e|-|i\rangle\langle i|$, and $L(A)=2A\rho A^{+}-A^{+}A\rho-\rho A^{+}A$ denote the Lindblad operators; $\kappa$, $\Gamma_{1}$, and $\Gamma_{2}$ are the decay rate of the cavity, the decay rate of the atom and the dephasing rate of the atom, respectively. Here, for a single qubit, we may choose $\kappa=0$, $\Gamma_{1}=\Gamma_{2}=2\pi\times2.6$ MHz, which has been experimentally demonstrated \cite{UQ,Atom,Spillane}. We consider the NOT and Hadamard gates as two typical examples, corresponding to $\varphi=\pi/4$ and $\varphi=\pi/8$ with the geometric phase $\gamma=\pi$. The maximum superadiabatic Rabi frequency $\Omega^{'}_{max}$ is a function of $\tau$. Moreover, one should guarantee that $\Omega^{'}_{max}$ is not larger than the peak Rabi frequency $\Omega_{max}$ of the original Hamiltonian $H(t)$. This constraint implies that we can only correct protocols with an effective protocol time $\tau>\tau_{min}\approx1/2.63\Omega_{max}$. For this reason, we set the effective Rabi strength of the driving field as $\Omega_{max}=2\pi\times750$ MHz, and $\tau=\tau_{min}$. For an initial state of the logical qubit $|\psi_{i}\rangle=|0\rangle$, the NOT and  Hadamard gates should result in the ideal final states $|\psi_{target_{N}}\rangle=|1\rangle$ and $|\psi_{target_{H}}\rangle=(|0\rangle-|1\rangle)/\sqrt{2}$, respectively.
The quality of the quantum gates is characterized by the fidelity $F=\langle\psi_{target}|\rho|\psi_{target}\rangle$. Numerical simulation of the fidelity dynamics for the superadiabatic transitionless
driving (SATD) and modified superadiabatic (MSA) single-qubit holonomic NOT and Hadamard gates are shown in Fig. 2(c) and Fig. 2(d), where the SATD  $F_N^S=98.66\%$  and MSA $F_N^M=99.10\%$ NOT gate and the SATD $F_H^S=99.55\%$ and MSA $F_H^M=99.70\%$ can be obtained, respectively. We also investigate the influence of the increase in the operation time and excited level decay on the gate fidelity, as shown in Fig. 2(e) and Fig. 2(f). We note that in Fig. 2(e) $F_H^A$ is defined as the fidelity of adiabatic Hadamard gate. Clearly, we find that the superadiabatic quantum gate can speed up the implementation of the adiabatic quantum gate in Fig. 2(c). Moreover, the superadiabatic quantum gate is more robust against decay, as shown in Fig. 2(d); for $\Gamma^{'}_{1}=20\Gamma_{1}$, we can still obtain the MSA gate $F^{M}_{N}\simeq93.8\%$ .


Next, we turn to the two-qubit controlled-phase (CP) gate. A holonomic two-qubit gate can be realized by controlling suitable coupling parameters between two three-level large detuning systems. To implement the scheme, we consider an atom-cavity architecture with two atoms trapped inside a cavity, as shown in Fig. {\ref{fig1}(c), with the level structure in (d). The coupling between the two atoms is mediated by a cavity in the Raman resonant regime. Therefore, when the Raman  resonant condition is satisfied, the effective Hamiltonian of the two-qubit gate can be taken as \cite{sm}
\begin{equation}\label{4}
 H_{1}(t)=\sum_{j=1}^{2}G_{j}(t)(\sigma_{j}^{-}a^{+}+\sigma_{j}^{+}a),
\end{equation}
where $G_j=\lambda_c\Omega_{R_j}/\Delta$ and the corresponding parameters are defined as $G(t)=\sqrt{G^{2}_{1}(t)+G^{2}_{2}(t)}$ with $G_{1}(t)=G(t)\sin\eta(t)$ and $G_{2}(t)=G(t)\cos\eta(t)$.

Similar to the single-qubit case, we take the Vitanov-style pulses as
\begin{eqnarray}
 \eta=\left\{
\begin{array}{ll}
\frac{\pi}{2+2e^{-(t-T/2)/\tau}},\ \ (0\leq t<T)\\
\frac{\pi}{2}- \frac{\pi} {2+2e^{-(t-3T/2)/\tau}},\ \ (T\leq t<2T)\\
\end{array}
\right.\notag\\
 G(t)=G_{max}.\ \ \ \ \ \ \ \ \ \ \ \ \ \ \ \ \ \ \ \  \ \ \ \ \ \ \ \ \ \ \ \ \ \ \ \ \ \ \  \ \ \ \ \ \ \ \ \ \ \ \
\end{eqnarray}
Under this condition,  the system is restricted in the subspace
 $\{|001\rangle, |100\rangle, |010\rangle, |110\rangle\}$, 
where $|001\rangle=|0\rangle_{1}\otimes|0\rangle_{2}\otimes|1\rangle_{c}$, i.e., they denote the states of the first and second atom and the cavity.
We can obtain the superadiabatic Hamiltonian to realize the superadiabatic holonomic two-qubit CP gate in the computational basis as
\begin{equation}\label{4}
 H_{2}(t)=G^{'}_{1}(t)|00\rangle\langle 10|+G^{'}_{2}(t)|00\rangle\langle 01|+\text{h.c.},
\end{equation}
where
\begin{eqnarray}\label{14}
\nonumber
\mu(t) &=&  -\arctan[\dot{\eta}(t)/G(t)],\\
\nonumber
\eta^{'}(t) &=& \eta(t)-\arctan[\dot{\mu}(t)/G(t)], \\
G^{'}(t) &=& \sqrt{G^{2}(t)+\dot{\mu}^{2}(t)},
\end{eqnarray}
with the control parameters $\eta$, undergoing a cyclic adiabatic evolution from $\eta(0)=0$. During the evolution, the $|10\rangle$ component adiabatically follows the dark state as  $\cos\eta(t)|10\rangle-\sin\eta(t)|01\rangle$, which acquires a Berry phase after the loop by the state $|10\rangle$, while the other states remain unchanged.
\begin{figure}[tbp]
\begin{center}
\includegraphics[width=0.98\columnwidth]{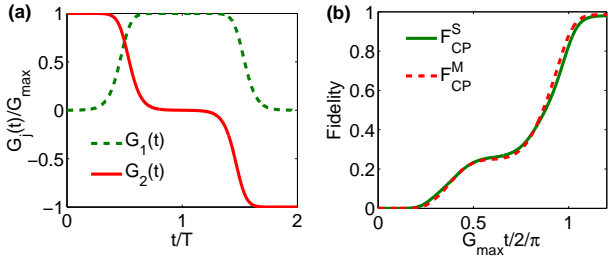}
\caption{(a) The original pulses for the superadiabatic two-qubit CP gate. (b) Two-qubit-state fidelity dynamics of
the superadiabatic two-qubit CP gate.}\label{fig3}
\end{center}
\end{figure}

Thus, we obtain the CP gate with the purely geometric phase $\gamma=\pi$. According to the original pulses, we can obtain the modified pulses by Eq. (\ref{14}) for realizing the two-qubit superadiabatic CP gate. Using the same approach of realizing the single-qubit gate, from $t=0$ to $t=T$, the Rabi frequency of the original Hamiltonian $H_{s}(0,T)$ with a constant phase $\phi=\phi_{1}=0$, and from $t=1T$ to $t=2T$, the Rabi frequency of the original Hamiltonian $H_{s}(T,2T)$ with a constant phase $\phi=\phi_{1}=\pi$. Under the cyclic evolution, the pure geometric phase $\gamma_{1}=\pi$ can be obtained. Therefore, we can achieve the following superadiabatic two-qubit CP gate:
\begin{equation}\label{cp}
U_{CP}=|0\rangle\langle0|\otimes\sigma_{z}+|1\rangle\langle1|\otimes\emph{I},
\end{equation}
where $\sigma_{z}$ is the Pauli matrix and \emph{I} denotes the unit $2\times2$ matrix.

We further verify the performance of the two-qubit CP gate. Under recent experimental conditions \cite{UQ,Atom,Spillane}, it has been predicted that the parameters $\lambda_{c}=2\pi\times750$ MHz, $\kappa=2\pi\times3.5$ MHz, and $\Gamma_{1}=2\pi\times2.6$ MHz, with the optical cavity mode wavelength in a range between 630 and 850 nm, can be achieved. We set $\Omega_{R}=\lambda_{c}$, $\tau=\tau_{min}$ and $T=2\tau$ and consider the driving field and cavity resonance having the same large detuning as $\Delta=2\pi\times4$ GHz.
For the initial state $|\psi_{i}\rangle=(|10\rangle+|G\rangle)/\sqrt{2}$, here, $|G\rangle=|000\rangle$, and the two-qubit CP gate should result in the ideal final states  $|\psi_{i}\rangle=(-|10\rangle+|G\rangle)/\sqrt{2}$. The fidelity of SATD and MSA two-qubit CP gates can reach approximately 97.99\% and 98.58\%, respectively, as shown in Fig. \ref{fig3}(b). Therefore, our robust scheme represents a feasible physical implementation with the strong coupling cavity QED.

\emph{Conclusion}. We have proposed a general scheme to realize universal superadiabatic holonomic quantum gates, with application to the C-QED system as a typical example. The designed universal gates are based on non-Abelian geometric phases, which can be robust against errors such as the decay of the open system by the master equation. The evolutions are superadiabatic
and thus can be fast and robust. The physical implementation of the scheme can be realized in the C-QED system with current technology.

This work was supported in part by the NFRPC (No. 2013CB921804), the NSFC (No. 11175067), the PCSIRT (No. IRT1243),
the NSF of Guangdong Province, China (No. S2011010003323), and the Program for Excellent Talents at the University of Guangdong Province.

\end{document}


\title{Supplemental Material for\\ "Superadiabatic Holonomic Quantum Computation in cavity QED''}
\author{Bao-Jie Liu}
\affiliation{Guangdong Provincial Key Laboratory of Quantum Engineering and Quantum Materials and School of Physics\\ and Telecommunication Engineering, South China Normal University, Guangzhou 510006, China}

\author{Zhen-Hua Huang}
\affiliation{Guangdong Provincial Key Laboratory of Quantum Engineering and Quantum Materials and School of Physics\\ and Telecommunication Engineering, South China Normal University, Guangzhou 510006, China}

\author{Zheng-Yuan Xue}\email{zyxue@scnu.edu.cn}
\affiliation{Guangdong Provincial Key Laboratory of Quantum Engineering and Quantum Materials and School of Physics\\ and Telecommunication Engineering, South China Normal University, Guangzhou 510006, China}

\author{Xin-Ding Zhang}\email{xdzhang@scnu.edu.cn}
\affiliation{Guangdong Provincial Key Laboratory of Quantum Engineering and Quantum Materials and School of Physics\\ and Telecommunication Engineering, South China Normal University, Guangzhou 510006, China}

\date{\today}
\maketitle

\subsection{Superadiabatic control for four-level system}
\indent We consider a tripod configuration in the scheme. The Hamiltonian is written as Eq.1 of the maintext. In the adiabatic basis, the Hamiltonian can be reduced to three-level form when $\varphi$ is kept constant. The reduced Hamiltonian in the bases $\{|+\rangle, |d_{1}\rangle, |-\rangle\}$ is Eq. (3) of the maintext, i.e.,
\begin{equation}\label{s1}
    H_{ad}(t)=\frac{\Omega(t)}{2}M_{z}+\dot{\theta}(t)M_{y},\\
\end{equation}
where the term $\dot{\theta}$   corresponds to the nonadiabatic couplings. In order to correct the nonadiabatic errors, there are two approaches.
Here, we modify the pulses via dressed states methods \cite{b1}. We look for a correction Hamiltonian $H_{C}(t)$ such that the superadiabatic Hamiltonian $H^{'}(t)=H(t)+H_{C}(t)$ governs a perfect state transfer. We choose the general form  
\begin{equation}\label{s1}
    H_{C}(t)=U^{+}(t)[g_{x}(t)M_{x}+g_{z}(t)M_{z}]U(t),\\
\end{equation}
where the unitary operator is defined, in the basis of \{$(\sin \varphi\left|0\right\rangle+\cos \varphi \left| 1 \right\rangle)$, $|e\rangle$, $|2\rangle$\}, as
\begin{eqnarray}
U(t)=\left(\begin{array}{cccc}
\sin\theta(t)/\sqrt{2} & -1/\sqrt{2} & \cos\theta(t)/\sqrt{2} \\
\cos\theta(t) & 0 & \sin\theta(t)\\
\sin\theta(t)/\sqrt{2} & 1/\sqrt{2} & \cos\theta(t)/\sqrt{2} \\
\end{array}\right),
\end{eqnarray}
Substituting Eq. 2 into $H_{C}(t)$, we get
\begin{equation}\label{1}
H^{'}(t)=\hbar[\Omega^{'}_{0}(t)|e\rangle\langle0|
+\Omega^{'}_{1}(t)|e\rangle\langle1|+\Omega^{'}_{2}(t)|e\rangle\langle a|]+H.c.,
\end{equation}
where the modified pulses are
\begin{eqnarray}
\Omega^{'}_{0} &=& (g_{z}(t)+\Omega(t))\sin\theta(t)\sin\varphi-g_{x}(t)\cos\theta(t)\sin\varphi, \nonumber\\
\Omega^{'}_{1} &=& (g_{z}(t)+\Omega(t))\sin\theta(t)\cos\varphi-g_{x}(t)\cos\theta(t)\cos\varphi,\nonumber\\
 \Omega^{'}_{a} &=& [(g_{z}(t)+\Omega(t))\cos\theta(t)+g_{x}(t)\sin\theta(t)]e^{-i\phi}.
\end{eqnarray}
Similar with the treatments in Ref. \cite{b1}, we define a new basis of dressed states  by the action of a time-dependent unitary operator $V(t)$ on the time-independent
eigenstates $|\pm,d_{1}\rangle$. In our model, without loss of generality, we takes $V(t)=e^{i\mu(t)M_{x}}$, with a Euler angle $\mu(t)$, moving in the frame defined by V in which the Hamiltonian
takes the form
\begin{equation}\label{1}
H_{new}(t)=VH_{ad}(t)V^{+}+VU(t)H_{C}(t)U^{+}(t)V^{+}+i\frac{dV}{dt}V^{+},
\end{equation}
For a perfect state transfer in new dressed states,the Hamiltonian $H_{new}(t)$ should be diagonalized. In other words, $H_{C}(t)$ has to
be designed such that  the unwanted off-diagonal elements in $H_{new}(t)$ equal zero. In order to satisfy Eq. (6) the control parameters
have to be chosen as
\begin{equation}\label{1}
g_{x}(t)=\dot{\mu(t)},  \quad g_{z}(t)=-\Omega-\frac{\dot{\theta}}{\tan{\mu(t)}}.
\end{equation}
The additional control Hamiltonian and dressed state basis must be chosen so that the dressed medium states coincide with the medium states at initial time and final time. Because of this condition, $\mu$ has to satisfy $\mu(t_{i})=\mu(t_{f})=0(2\pi)$. The simplest nontrivial choice of the dressed states basis is the superadiabatic basis, for which
\begin{equation}\label{1}
\mu=-\arctan\left(\frac{\dot{\theta}}{\Omega(t)}\right), \quad g_{x}=\dot{\mu}, \quad g_{z}=0,
\end{equation}
This choice will be referred to as superadiabatic transitionless driving (SATD). To reduce the intermediate-level occupancy, we can further modify the SATD by choosing the suit parameter as \cite{b1}
\begin{eqnarray}\label{1}
\mu&=&-\arctan\left[\frac{\dot{\theta}}{f(t)\Omega(t)}\right],\notag\\
g_{x}&=&\dot{\mu},\notag\\
g_{z}&=&-\Omega-\frac{\dot{\theta}}{\tan{\mu(t)}}.
\end{eqnarray}
Taking Eq. (8) into Eq. (5) we can get the modified diving field strength with SATD as Eq. (4) of the maintext.
Similarly, taking Eq. (8) into Eq. (5), we can obtained similar result with modified superadiabatic transitionless driving (MSA). We plot the modified pulses for realizing  NOT gate with (a) and (d), Hardarmd gate with (b) and (e), and two-qubit control phase gate with (c) and (f) by the SATD and MSA in the Fig. 1, and these modified pules can achieve high fidelity quantum gate.

\begin{figure*}[tbp]
\begin{center}
\includegraphics[width=0.9\textwidth]{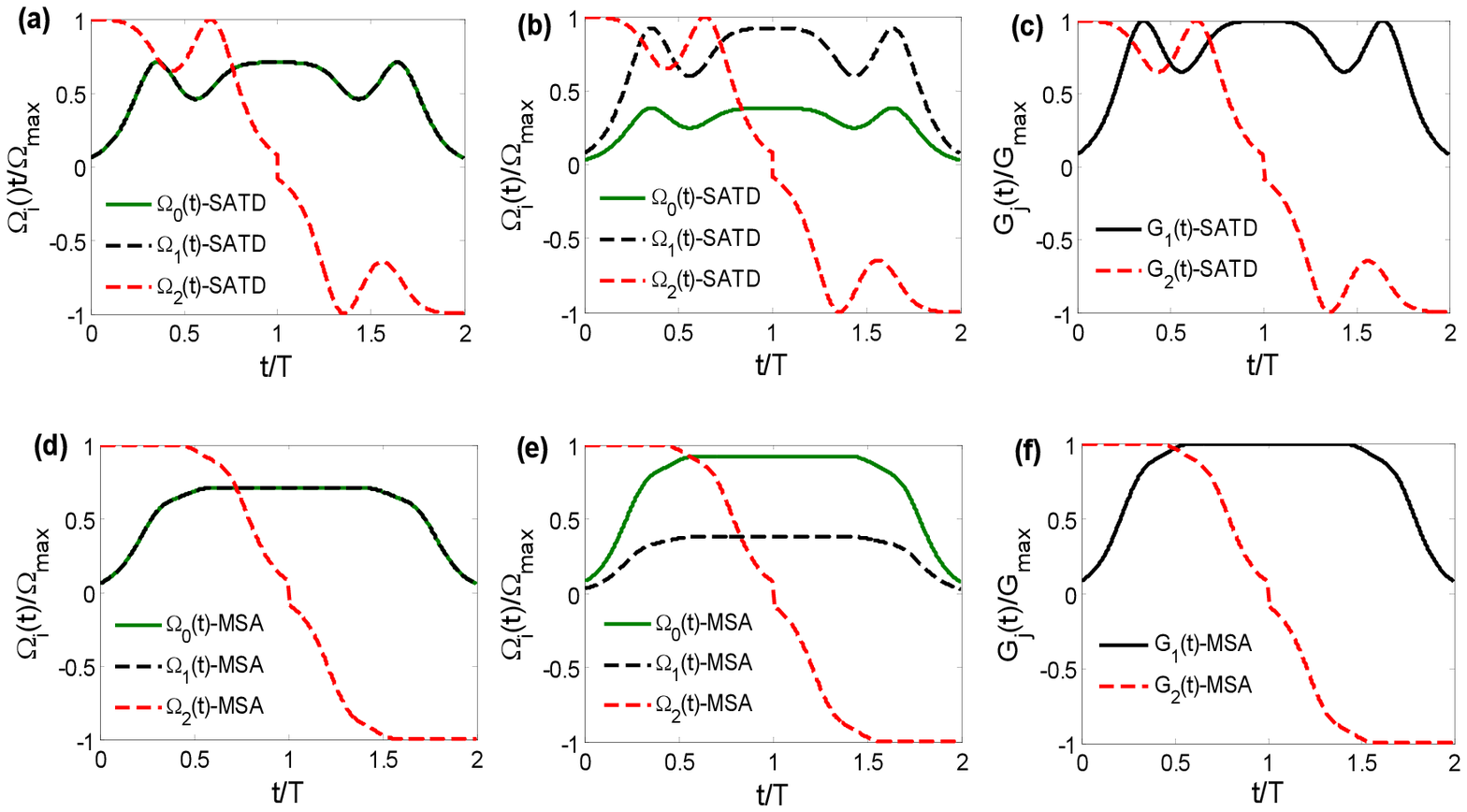}
\caption{Schematic diagram of superadiabatic pulses (SATD) and modified superadiabatic pulses (MSA) for realizing  the NOT gate with (a) and (d),Hardarmd gate with (b) and (e), and two-qubit control phase gate with (c) and (f).}\label{1}
\end{center}
\end{figure*}

\subsection{Effective atom-cavity coupling}
Our scheme relies on the Raman excitation of two three-level atom by a driving field of frequency $\omega_{R_{j}}$ and a quantized cavity mode of frequency $\omega_{0}$ in a lambda configuration. The field drives dispersively the transition from level $6^2S_{1/2}$, $|1\rangle=|F=4,m=3\rangle$ to level $6^2P_{1/2}$, $|e\rangle=|F=4,m=3\rangle$, with coupling strength $\Omega_{R}$ and detuning $\Delta=\omega_{e1}-\omega_{R}\gg |\Omega_{R_{j}}|$. The cavity mode couples level $6^2S_{1/2}$,$|0\rangle=|F=3,m=2\rangle$ to level $6^2P_{1/2}$ ,$|e\rangle=|F=4,m=3\rangle$, with coupling constant $\lambda_{c}$ and the same detuning $\Delta=\omega_{e0}-\omega_{R}\gg |\lambda_{c}|$. In the interaction picture, the interaction Hamiltonian in the rotating wave approximation is  
 \begin{equation}\label{1}
H_{int}=\sum_{j}(\Omega_{R_{j}}\sigma^{j}_{e1}e^{-i\Delta t}+\lambda_{c}\sigma^{j}_{e0}ae^{-i\Delta t}+\text{h.c.})
\end{equation}
where $\sigma_{km}=|k\rangle\langle m|$ is an electronic flip operator, and $a$ ($a^{+}$) is the annihilation (creation) operators of the quantized cavity mode. For sufficiently large
$\Delta$, negligible population is transferred to $|e\rangle$. When the Raman resonance condition   also holds, Eq. (14)of the maintext  can be rewritten as
\begin{equation}\label{N}
 H_{1}(t)=\sum_{j=1}^{2}G_{j}(t)(\sigma_{j}^{-}a^{+}+\sigma_{j}^{+}a),
\end{equation}
where $\sigma_{-}=|1\rangle\langle 0|$ and the effective coupling strength is $G_{j}=\frac{\lambda_{c}\Omega_{R_{j}}}{2\Delta}$.

